\documentstyle[preprint,aps,tighten,prb,epsf]{revtex}
\begin{document}
\draft
\preprint{(Nordita preprint $\#$ 96/9, submitted to Phys. Rev. B)}
\title{Laser-induced quantum chaos in 1-D crystals} 
\author{H. S. Brandi, Belita Koiller}
\address{Instituto de F\'\i sica, Universidade Federal do Rio de Janeiro, 
Cx. P. 68.528, 21945-970, Rio de Janeiro, Brazil}
\author{and}
\author{Eduardo R. Mucciolo}
\address{NORDITA, Blegdamsvej 17, DK-2100 Copenhagen {\O}, Denmark}
\date{February 9, 1996}
\maketitle
\begin{abstract}
We study the electronic band structure for a model one-dimensional
periodic potential in the presence of a spacially homogeneous laser
field. The statistical properties of the energy bands depend on the
coupling between the crystal and the laser field, going from Poisson
to Wigner-Dyson (GOE) and back to Poisson as the coupling
increases. We argue that the classical dynamics of this system
resembles that of a periodically driven pendulum. We find that the
chaotic regime is not restricted solely to high-lying bands and should
thus be of easier access to optical experiments.
\end{abstract}
\narrowtext

In the past decade, many model systems exhibiting what conventionally
became known as quantum chaos have been discussed in the
literature.\cite{gutzwiller90,haake91,leshouches91} Among those, the
kicked rotator\cite{casati79,moore94,robinson95,bardroff95} and the
Hydrogen atom in the presence of a uniform magnetic field \cite{hatom}
have been studied in great detail, both theoretically and
experimentally. Formally, one speaks of quantum chaotic behavior when
the system dynamics in the classical limit is very irregular or
completely chaotic. For a quantum system whose classical dynamics is
regular one usually finds that energy levels are uncorrelated and
therefore obey a Poissonian statistics. On the other hand, if the
system is fully chaotic in the classical limit, its energy levels tend
to repel each other strongly. In this case the statistical properties
of the spectrum coincide with those drawn from the Gaussian ensembles
of random matrix theory (RMT).\cite{RMT}

Recently, Mucciolo and coworkers showed\cite{mucciolo94} that the
electronic spectra of crystalline solids (e.g., Si) also show the
universal signatures of quantum chaotic behavior. Their analysis was
restricted to high-lying bands, well above the Fermi energy. Since
crystals are fixed by nature, no external parameter exists to modify
the system and therefore try to enhance the irregularity in its
classical dynamics. This is in contrast with the two previously
mentioned systems,\cite{casati79,moore94,robinson95,bardroff95,hatom}
where chaos is controlled by an external parameter, as for example the
strength of a time-periodic impulse or the intensity of a magnetic
field.

In this work we present a new model exhibiting spectral signatures of
chaos based on electrons in a periodic potential. In our case a
monochromatic laser field is the main responsible for the irregular
electron dynamics. The intensity of the laser field can be regarded as
an external parameter changing the behavior of the system. We believe
that this mechanism could increase the chances of experimentally
observing chaos in crystals by bringing universal statistics to
low-lying bands as well. By looking at the energy level statistics of
a very simple 1-D periodic potential we were able to identify three
regimes. These suggest that the system dynamics moves from regular to
chaotic and then back to regular as the field intensity is
increased. Our formalism is based on the dressed band
approach,\cite{jalbert86} which we briefly review below.

According to Bloch's theorem, the wave function for an electron in a
1-D periodic potential can be labelled by a momentum $k$, $-\pi/d\leq
k<\pi/d$, where $d$ is the lattice parameter. This allows the
Hamiltonian to be decoupled into well-defined $k$ (reduced)
components. Inclusion of the interaction with a monochromatic laser
field leads to the following reduced Hamiltonian:
\begin{equation}
H=H_k + H_\gamma + H_{int} \;,
\label{eq:tothamilt}
\end{equation}
where, in atomic units ($\hbar = e = m = 1$), 
\begin{eqnarray}
H_k & = & {p^2\over 2} + V(z) \nonumber \\ & = & -{1\over 2} \left(
{d\over {dz}} + ik
\right)^2 + \sum_\ell V(G_\ell) \exp(i\,G_\ell z) \; ,
\label{eq:elechamilt}
\end{eqnarray}
\begin{equation}
H_\gamma = \omega\,a^+a \;,
\label{eq:phothamilt}
\end{equation}
and
\begin{equation}
H_{int} = \alpha \, Ap + {\alpha ^2 \over 2} A^2 \;.
\label{eq:interhamilt}
\end{equation}
In the above equations, $V(z)$ denotes the lattice potential, $a^+$
$(a)$ is the creation (annihilation) field operator, $\alpha = 1/137$
is the fine structure constant, and $A = {1\over\alpha}
\sqrt{2\pi/\omega\Omega}\, (a+a^+)$, where $\Omega$ is the
quantization volume. The periodicity of the lattice potential allows
it to be written in terms of a reciprocal lattice sum over
$G_\ell=2\pi\ell/d$, as indicated in Eq.~(\ref{eq:elechamilt}). Here
the field polarization is taken along the $z$-direction.

By performing adequate unitary transformations in terms of a basis of
eigenstates $|G,n \rangle$ of the momentum $p$ and the field
Hamiltonian $H_\gamma$, $H$ can be transformed into a Floquet
Hamiltonian $H_F$. One can then show\cite{jalbert86} that the spectrum
of $H$ follows from the diagonalization of the Bloch-Floquet matrix
\begin{equation}
\langle G',n'|H_F - N_0\omega |G,n\rangle = \left[ n\omega + 
{(G+k)^2 \over 2} \right] \delta_{G,G'}\delta_{n,n'} + J_{n'-n}\!\left(
{\alpha A_0 \over\omega} (G-G')\right) V_{G-G'} \;,
\label{eq:blochfloquet}
\end{equation}
where $A_0=\frac{2}{\alpha}\sqrt{2\pi N_0/\omega\Omega}$ is the
amplitude of the classical vector potential and $J_n$ is the Bessel
function of order $n$. It should be noted that $n$ in the state
$|G,n\rangle$ denotes an integer (positive or negative) that describes
the deviation from the average number of photons $N_0$, which is
supposed to be large. Note that the matrix defined by
Eq. (\ref{eq:blochfloquet}) is time-reversal symmetric.

We present results for a model potential where $V(G_\ell) = -\sigma$
for $-3\le\ell\le 3$ and zero for other $\ell$ values. This is a
truncated Kronig-Penney attractive model. We take the energy unit to
be $\epsilon = G_1^2 / 2$ and the potential strength $\sigma = 0.25
\epsilon$. The laser intensity is characterized by the dimensionless
parameter $x=\alpha A_0 G_1/\omega$. Equation~(\ref{eq:blochfloquet})
allows for a simple interpretation of the spectrum in this problem: In
the absence of the laser field, the electronic energy levels consist
essentially of the free-electron parabola folded into the first
Brillouin zone (BZ), plus gaps opening around the crossing points
($k=0$ and $ \pm G_1/2$). The gap amplitude and the distortion of the
bands with respect to the free-electron parabola depend on the
potential strength $\sigma$. Inclusion of the field produces a
dressing effect in the bands, which may be described by replica-bands
translated by an amount $n\omega$. The interaction $H_{int}$ causes an
anti-crossing whenever two noninteracting replica-bands
cross. Clearly, this occurs for all $k$ in the BZ; for a fixed value
of $\omega$, the gaps widen as the intensity $x$ increases.

\begin{figure}
\setlength{\unitlength}{1mm}
\begin{picture}(150,170)(0,0)
\put(3,10){\epsfxsize=15cm\epsfbox{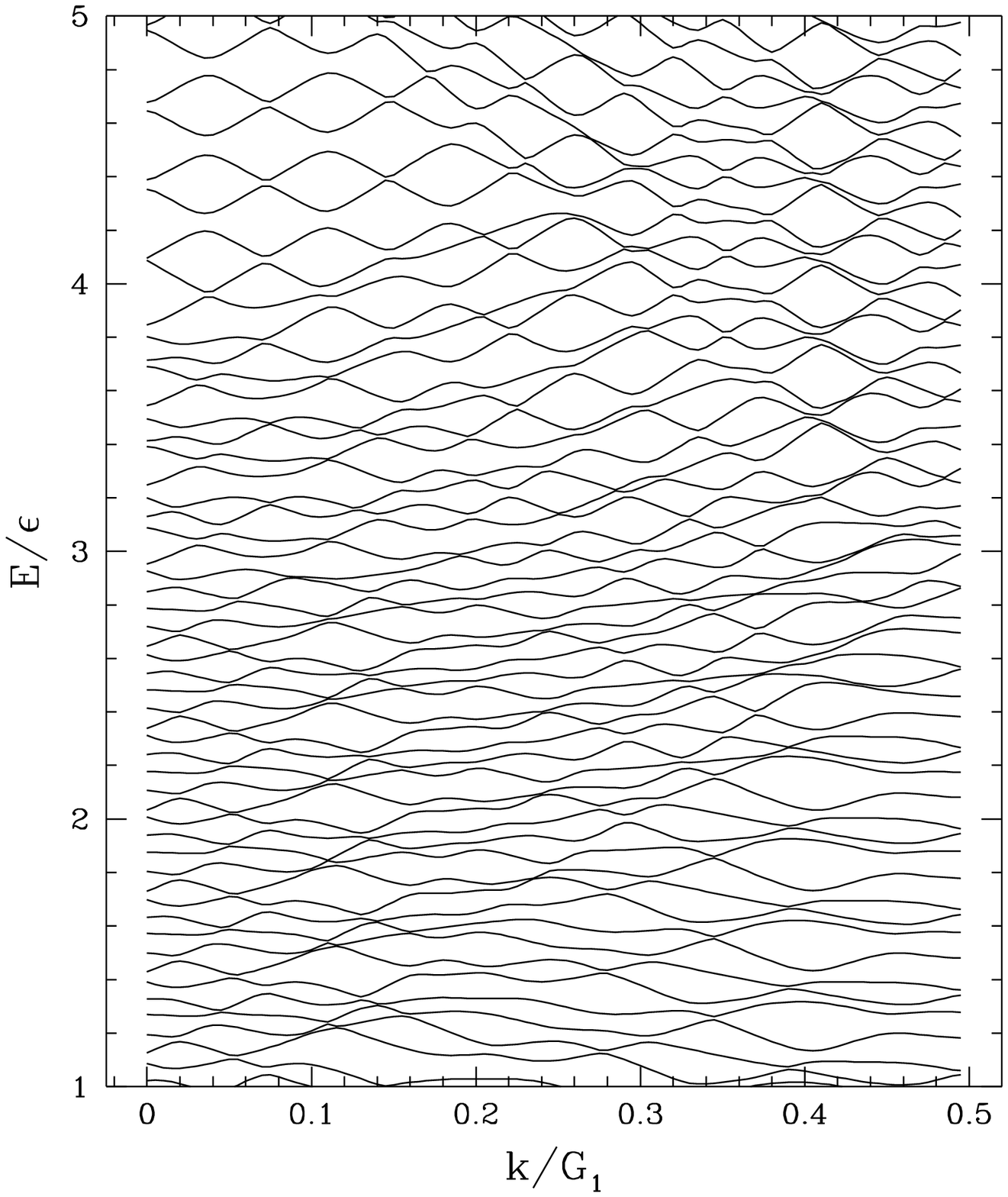}}
\end{picture}
\caption{The electronic spectrum in a 1-D periodic potential under a laser
field of intensity $x=10$. Energy bands are shown as a function of the
Bloch momentum $k$.}
\label{fig:spectra}
\end{figure}

In Fig.~\ref{fig:spectra} we show the spectrum obtained for $-10 \le
n\le 10$, $\omega = 0.3 \epsilon$, and field intensity $x = 10$. We
avoid regions at the extremes of the energy spectrum, where only
replicas of the lowest or higher bands are present (due to the
truncation in $n$ adopted in our numerical treatment) and limit our
analysis to the energy region $1<E/\epsilon <5$, where a large number
of anti-crossings occur. For the parameters chosen here, this energy
region is representative of the complete basis, $-\infty < n
<\infty$. The laser frequency $\omega$ is chosen to maximize the
mixing between the dressed bands. Note that the spectrum in
Fig.~\ref{fig:spectra} reveals an intricate level repulsion structure
which resembles those characteristic of quantum chaotic regimes in
other systems.\cite{haake91,leshouches91,hatom,mucciolo94}

A quantitative identification of quantum chaos consists primarily in
unfolding the energy spectrum (to attain a constant density of states)
and then analyzing the short and long-range statistical correlations
present. For spinless, time-reversal symmetric systems with a complex
dynamics, RMT\cite{RMT} predicts that the distribution of
nearest-neighbor level spacings (NNS) should follow closely the Wigner
surmise for the Gaussian orthogonal ensemble (GOE),
\begin{equation}
P^{\rm GOE} (s) = \frac{\pi}{2}s\ e^{-\pi s^2/4} \;,
\label{eq:levspacGOE}
\end{equation}
where the level spacing $s$ is given in units of the mean level
spacing. On the other hand, the energy levels of most integrable
systems are essentially uncorrelated (one important exception is the
harmonic oscillator), in which case the NNS distribution is a simple
exponential, namely,
\begin{equation}
P^{\rm Poisson} (s) = e^{-s} \;.
\label{eq:levspacPoi}
\end{equation}
Long-range correlations can be diagnosed through the least square
deviation $\Delta_3(L)$ of the number of levels in a given energy
interval of length $L$ from a constant slope.\cite{RMT} When $L$ is
given in units of the mean level spacing, $\Delta_3(L)$ is expected to
fall between two extreme limits: linear ($L/15$) for uncorrelated
levels and constant ($1/12$) for fully rigid, equally spaced
spectra. The GOE prediction\cite{RMT} is an intermediate function
which, for large $L$, behaves as $\Delta_3^{\rm GOE}(L) \approx
(1/\pi^2) \ln(2\pi L) - 0.007$.

\begin{figure}
\setlength{\unitlength}{1mm}
\begin{picture}(160,180)(0,0)
\put(0,5){\epsfxsize=16cm\epsfbox{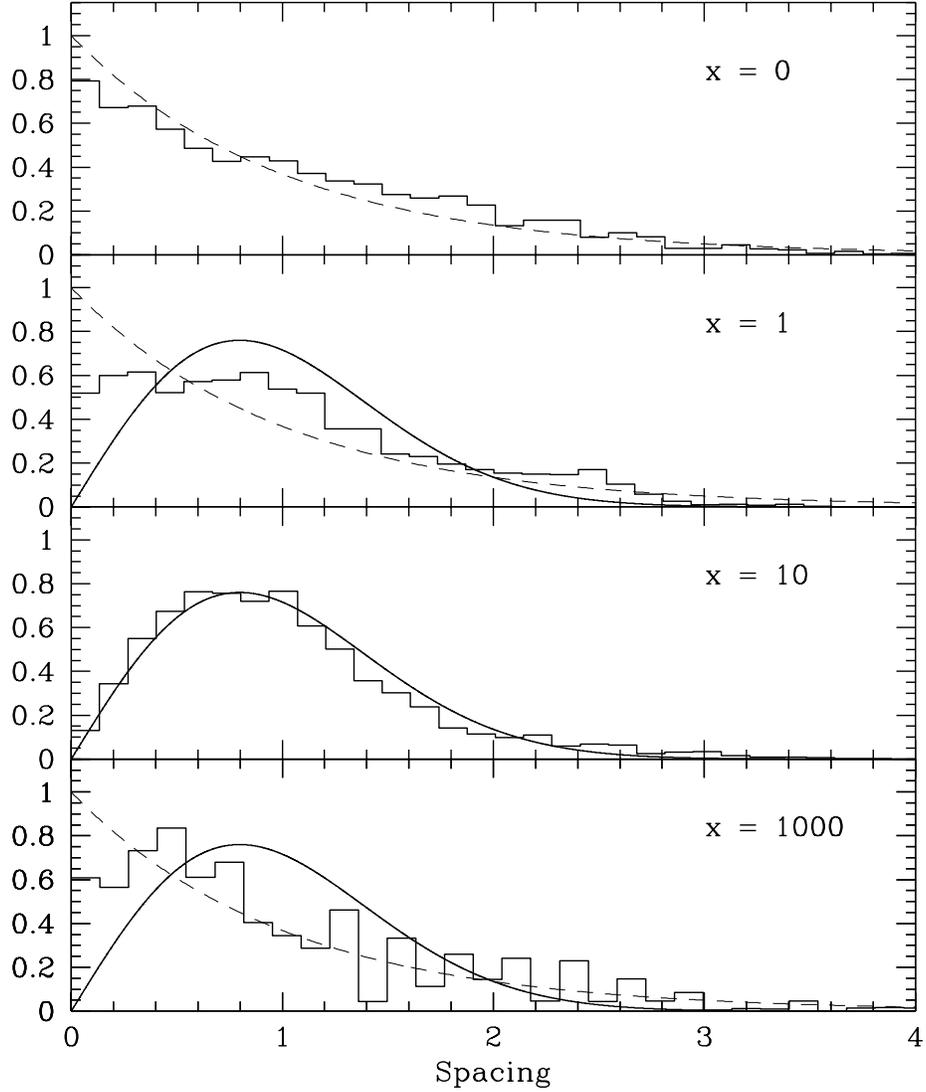}}
\end{picture}
\caption{Histograms giving the nearest-neighbor spacing distributions for 
the energy spectra of Eq.~(\protect{\ref{eq:tothamilt}}) under
increasing field intensity (from top to bottom). Solid and dashed
lines correspond to GOE and Poisson statistics respectively. Note
that, as the field intensity increases, the distribution evolves from
Poisson-like into GOE-like, and then back to Poisson-like.}
\label{fig:levspac}
\end{figure}

\begin{figure}
\setlength{\unitlength}{1mm}
\begin{picture}(150,150)(0,0)
\put(0,0){\epsfxsize=15cm\epsfbox{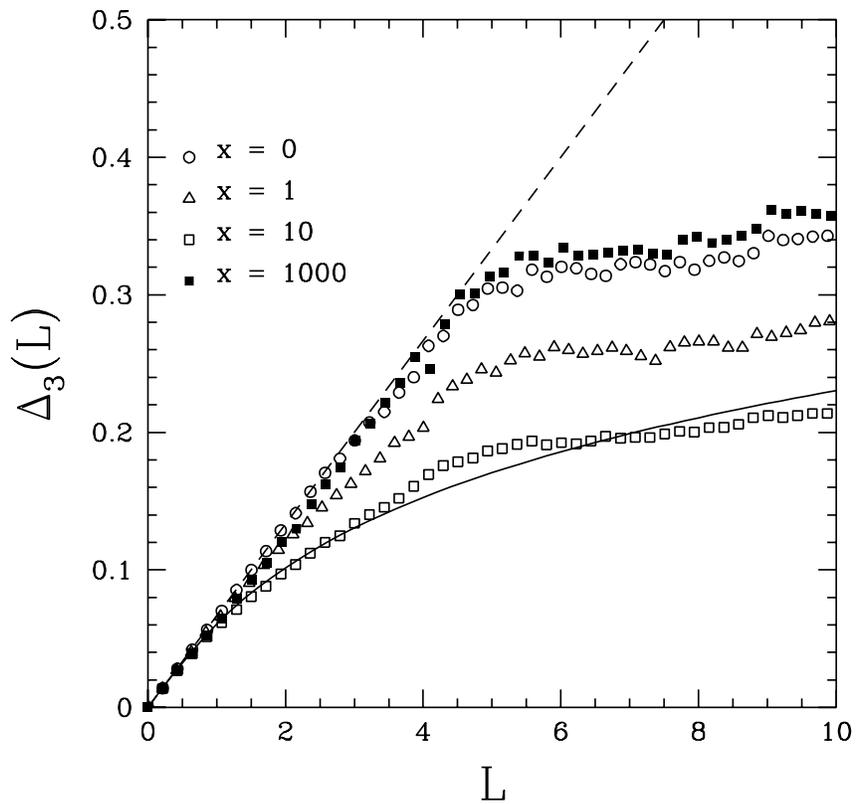}}
\end{picture}
\caption{Least square deviation statistics $\Delta_3(L)$ for the
indicated values of the field intensity parameter $x$. The solid and
dashed lines are the GOE and Poisson regimes, respectively.}
\label{fig:delta3}
\end{figure}

In Fig.~\ref{fig:levspac} and \ref{fig:delta3} we present our results
for the NNS distributions and $\Delta_3$, respectively, obtained for
increasing values of the field intensity parameter $x$. The data
correspond to a sampling over 100 equidistant $k$ values along $0\leq
k/G_1\leq 0.5$. The level density remained essentially constant on the
energy interval considered and no unfolding was carried out. The
resulting statistics involved over 5,200 calculated levels for each
$x$. Let us concentrate initially on Fig.~\ref{fig:levspac}. For $x=0$
the uncorrelated character of the energy levels and the integrability
of the Hamiltonian in Eq.~(\ref{eq:tothamilt}) is reflected in the
agreement between the data and the Poisson law. As $x$ increases, a
weak level repulsion appears. Note that $x=1$ corresponds to a
crossover regime in which neither of the distributions in
Eqs.~(\ref{eq:levspacPoi}) and (\ref{eq:levspacGOE}) agrees with the
data. However, for $x=10$ the Wigner surmise provides a very accurate
fit, suggesting a chaotic regime at this intensity. Further increase
in $x$ causes the data to change, deviating from the Wigner surmise.
For $x=1000$, the spectrum tends again to a Poisson statistics, which
reflects the exact integrability in the limit of infinite field
intensity as well. It is important to note that the NNS is not a
delta-function centered at $\omega$ when $x\rightarrow\infty$: The
periodic potential, although relatively weak, folds the sequence of
equidistant parabolas back onto the first BZ, causing the NNS to be
Poissonian.

A similar interpretation goes for Fig.~\ref{fig:delta3}. The maximal
spectral rigidity occurs around $x=10$. In this figure, however, we
see a marked deviation from either the Poisson ($x=0$ and 1000) or the
GOE ($x=10$) predictions beyond intervals of length $L_{\rm
max}\approx 4$. From Fig~\ref{fig:spectra} we may notice a tendency of
levels to form clusters of four. This effect is well known in the
literature:\cite{casati85,berry91} In a semiclassical language, it
indicates the breakdown of the universal regime and the appearance of
system-dependent contributions, which are related to very short
periodic orbits. We have checked that $L_{\rm max}$ is always
approximately equal to the laser frequency $\omega$ (in units of the
mean level spacing). The periodic orbit associated to this frequency
is the oscillation of the laser field in its phase space.\cite{maser}
The reason why this orbit is always the shortest possible one and its
signature survives different values of $x$ is related to the nature of
our model. For the Kronig-Penney model there is only one bound state
($E<0$) in the spectrum. As a result, in the energy range where the
$\Delta_3$ was obtained, the electron classical motion can only be
bounded (and therefore periodic) through the interaction with the
laser field. This always leads to a periodicity $T> 2\pi/\omega$ and
the intrinsic period of oscillation of the laser field remains the
shortest one.

As noted before, for a 3-D periodic potential the electronic spectrum
at low-symmetry regions of the BZ manifests quantum chaotic
behavior. In the equivalent 1-D problem, no chaotic behavior is
expected, since the electrons have only one degree of freedom $(z)$
and the problem is classically integrable. The introduction of a laser
field adds another degree of freedom to the system and makes the
problem nonintegrable. As indicated by our numerical results, for high
enough intensities of the laser field, it is plausible that this
perturbation causes the electron motion to be strongly irregular,
leading to chaos.

Although we treat the laser field in a second quantized form, our
results are equivalent to those obtained from a semiclassical,
time-dependent description where the field is fixed and can be written
in the form $A_0\cos (\omega t)$. Both descriptions yield the same
spectrum (band structure) of energies and quasi-energies in the
quantized and semi-classical descriptions, respectively. The only
difference is that, in the semiclassical case, one formally
incorporates the time-periodic vector potential to the crystal
potential, resulting in a space-time (two-dimensional) BZ, but with
matrix elements for the effective Hamiltonian similar to
Eq. (\ref{eq:blochfloquet}).

It is interesting to compare the problem of a laser-illuminated 1-D
crystal to the kicked rotator.\cite{haake91,casati79} The dynamics in
the latter is generated by a Hamiltonian in which the free rotator is
subject to a time-periodic force whose amplitude is a function of the
rotation angle ($2\pi$-periodic in $\theta$). In the classical
counterpart of our model (and in the kicked rotator), an external
field, space and time periodic, acts on a particle moving in one
dimension. The difference between these two systems is that, in our
case, space and time components are decoupled, allowing one of the
potentials to be made negligible with respect to the other by varying
the parameter $x$. Thus, the $x=0$ limit corresponds to an electron in
a 1-D crystal problem, and $x\to\infty$ corresponds to a free electron
in the presence of the laser field. In both limits the problem is
exactly integrable.

No such decoupling is possible in the kicked rotator. Our problem
would be more closely related to a periodically kicked {\it pendulum},
in which case the kick amplitude $\lambda$ would be a control
parameter similar to the laser intensity, and this system should
present quantum chaos for intermediate values of $\lambda$. For
$\lambda=0$ and for $\lambda\to\infty$ the problem would be dominated
by the gravitational field or by the kick force respectively, becoming
integrable and therefore non-chaotic.

To our knowledge, no other system discussed in the literature has been
led from regularity into chaos and than back to regularity by
increasing a single control parameter. One might think that a similar
situation would occur for the H atom in a magnetic field since the
problem is exactly integrable in the limits of zero magnetic field and
zero Coulomb potential. However, even in the limit of extremely strong
magnetic field, the Coulomb interaction can not be neglected. In this
limit the classical motion of the electron is strongly constrained in
the field direction and the electronic distribution is no longer
spherically symmetric. The electronic motion may be approximated by a
1-D truncated Coulomb interaction with the nucleus, which has a
completely different spectrum from the Landau problem.\cite{hatom}
Other simple model-Hamiltonians, such as the the spin-boson
model,\cite{caio91} may display a variety of behaviors, but they
require variations of {\it several} control parameters.

Finally, we discuss the experimental implications of our
model. Contrary to the 3-D band structure problem,\cite{mucciolo94}
the chaotic behavior presented here is not limited to highly excited
bands. Any region of the spectrum of the 1-D system displays the
universal features under a laser field of adequate intensity. The same
effect should be present in 3-D systems, even for bands close to the
Fermi energy. The limitation on the long-range universal correlations
imposed by short periodic orbits should become less stringent in
higher dimensions due to the increase in the density of states. We
believe that optical pump-probe techniques should be adequate to
investigate the dressing effects presented here. They should reveal
universal signatures in the frequency and momentum dependence of the
dielectric function and the optical conductivity, as predicted by
Taniguchi and Altshuler.\cite{nobuhiko93}

\acknowledgements
We thank H.~Chacham for useful discussions and C.~Lewenkopf for
interesting suggestions and a critical reading of the manuscript. This
work was supported in part by the Conselho Nacional de Desenvolvimento
Cient\'\i fico e Tecnol\'ogico (CNPq-Brazil).

\end{document}